\documentclass[usenatbib]{mn2e} 
\usepackage{amsmath} 
\usepackage{amssymb} 
\usepackage{graphics}
\usepackage{epsfig}
\def\be{\begin{equation}}
\def\ee{\end{equation}}
\def\ba{\begin{eqnarray}}
\def\ea{\end{eqnarray}}

\usepackage{color}
\definecolor{red}{rgb}{1,0.0,0.0}

\usepackage[normalem]{ulem}
\definecolor{darkgreen}{rgb}{0.0,0.5,0.0}

\newcommand{\beq}{\begin{eqnarray}}  
\newcommand{\eeq}{\end{eqnarray}}  
  
\newcommand{\apj}{ApJ}  
\newcommand{\apjs}{ApJS}  
\newcommand{\apjl}{ApJL}  
  
\newcommand{\mnras}{MNRAS}  
\newcommand{\mnrassub}{MNRAS accepted}  
\newcommand{\aap}{A\&A}  
  
\newcommand{\araa}{ARA\&A}

\newcommand{\avg}[1]{\langle{#1}\rangle}  
\newcommand{\ly}{{\ifmmode{{\rm Ly}\alpha}\else{Ly$\alpha$}\fi}}
\newcommand{\hMpc}{{\ifmmode{h^{-1}{\rm Mpc}}\else{$h^{-1}$Mpc }\fi}}  
\newcommand{\hGpc}{{\ifmmode{h^{-1}{\rm Gpc}}\else{$h^{-1}$Gpc }\fi}}  
\newcommand{\hmpc}{{\ifmmode{h^{-1}{\rm Mpc}}\else{$h^{-1}$Mpc }\fi}}  
\newcommand{\hkpc}{{\ifmmode{h^{-1}{\rm kpc}}\else{$h^{-1}$kpc }\fi}}  
\newcommand{\hMsun}{{\ifmmode{h^{-1}{\rm {M_{\odot}}}}\else{$h^{-1}{\rm{M_{\odot}}}$}\fi}}  
\newcommand{\hmsun}{{\ifmmode{h^{-1}{\rm {M_{\odot}}}}\else{$h^{-1}{\rm{M_{\odot}}}$}\fi}}  
\newcommand{\Msun}{{\ifmmode{{\rm {M_{\odot}}}}\else{${\rm{M_{\odot}}}$}\fi}}  
\newcommand{\msun}{{\ifmmode{{\rm {M_{\odot}}}}\else{${\rm{M_{\odot}}}$}\fi}}  
\newcommand{\lya}{{Lyman-$\alpha$ }}

\newcommand{\rand}{{\ifmmode{{\mathcal{R}}}\else{${\mathcal{R}}$ }\fi}}  
\begin{document}

\title[LAE/LBG ratio at high redshift]{Modelling the fraction
  of Lyman Break Galaxies with strong Lyman-$\alpha$ emission at $5\leq z \leq 7$}
\author[Forero-Romero et al.]{
\parbox[t]{\textwidth}{\raggedright 
  Jaime E. Forero-Romero$^1$ \thanks{Email: jforero@aip.de}, Gustavo Yepes$^2$, Stefan Gottl\"ober$^{1}$ \\
  \& Francisco Prada$^3$}\\
\vspace*{6pt}\\
$^1$Leibniz-Institut f\"ur Astrophysik Potsdam (AIP), An der Sternwarte 16, 14482 Potsdam, Germany\\ 
$^2$Grupo de Astrof\'{\i}sica, Universidad Aut\'onoma de Madrid,   Madrid
E-28049, Spain\\
$^3$Instituto de Astrof\'{\i}sica de Andaluc\'{\i}a (CSIC), Camino Bajo de Hu\'etor 50, E-18008, Granada, Spain
}
\maketitle

\begin{abstract}
  We present theoretical results for the expected fraction of
  Lyman Break Galaxies (LBGs) to be detected as strong Lyman-$\alpha$ emitters
  (LAEs) in the   redshift range $5\leq z \leq 7$. We base our analysis on the
  2-billion particle SPH simulation {\it MareNostrum High-z Universe}. 
  We approximate galaxies as static dusty slabs with an additional
  clumpy dust distribution affecting stellar populations younger than $25$ Myr.
  The model for the Lyman-$\alpha$ escape fraction is based on the results of
  our Monte-Carlo radiative transfer code ({\tt CLARA}) for a slab
  configuration.  We also fix the transmission of \lya photons through the intergalactic medium to a constant value of $50\%$ at all redshifts. From the results of this model we calculate $x_{\ly}$, the fraction of Lyman
  Break Galaxies with \ly\ equivalent width (EW) larger than $50$\AA. We find a remarkable agreement with observational data at
  $4.5<z<6$. For bright ($-22 <M_{\rm UV}<-20.5$) and faint ($-20.5<M_{\rm
  UV}<-18.5$) galaxies our model predicts $x_{\ly}=0.02\pm 0.01$ and
  $x_{\ly}=0.47\pm 0.01$ while observers report $x_{\ly}=0.08\pm0.02$ and
  $x_{\ly}=0.47\pm0.16$, respectively.  Additional evolution of the extinction model at redshift $z\sim 7$, that decreases the intensity of transmitted \lya radiation  by a factor of $f_{\rm T}=0.4$ as to match the LAE luminosity function  at $z\sim 6.5$, naturally  provides a good match for the recently reported $x_{\ly}$  fractions at  $z> 6.3$. Exploring different toy models for the
  \lya\ escape fraction, we show that a decreasing \lya escape fraction with
  increasing UV galaxy luminosity is a key element in our model to explain the trend  of larger $x_{\ly}$ fractions for fainter LBGs. 
\end{abstract}


\section{Introduction}

The study of distant star forming galaxies is
being driven by even more sensitive observations. Observational samples are
now commonly gathered both for Lyman Break Galaxies (LBG) and Lyman-$\alpha$
emitting (LAE) galaxies in the redshift range $z>4$
(see \cite{2011ApJ...730....8H} and references therein). Nevertheless, a clear
physical connection between these two populations is difficult to establish
given the complex physics involved in the transmission Lyman-$\alpha$
radiation in the interstellar medium (ISM) \citep{2006MNRAS.367..979H}. 

Recently, observational results have reported on the fraction of LBGs that
show strong \lya\ emission, $x_{\ly}$, at a given absolute rest-frame UV
magnitude \citep{stark1,stark2,2011arXiv1107.1261S}. This fraction derived from observational data
provides a simple test to theoretical models that seek to explain, under a sound
physical model, the connection between LBGs and LAEs. Because this test is
concerned with fractions of a population, it is largely insensitive to changes
in absolute number densities of the observed galaxies, adding new information
with respect to analysis based on the LAE luminosity function (LF). 

In this paper we make theoretical predictions on the $x_{\ly}$
fraction as function of absolute rest-frame magnitude $M_{\rm UV}$ in the
redshift range $5\leq z\leq7$. These results are based on a simulation of a
cosmological volume that follows the dynamical evolution of roughly 2 billion
dark matter and gas particles including star formation and supernovae
feedback. The clumpy dust attenuation model and its effects on the UV
luminosity function were described in detail in \cite{UVletter} (Paper I hereafter). The
corresponding \lya\ escape fraction is calculated using the results of our
Monte-Carlo code {\tt CLARA}. In \cite{CLARA} (Paper II hereafter) we
discussed at length the structure of the code and the implications for the
luminosity functions in the redshift range $5\leq z \leq7$.  

In Section \ref{sec:galfinder} we describe the numerical simulation and the
model for the UV/\lya\ emission and its associated extinction/escape
fraction. Next (\S \ref{sec:results})we describe the results of this model for
the $x_{\ly}$  statistic in the redshift range $5\leq z\leq 7$. We also
explore different toy models for the Lyman $\alpha$ escape fraction in order
to better understand the trends found in the simulation. We discuss the results
in Section \ref{sec:discussion} and present our conclusions in Section
\ref{sec:discuss}.

\section{A Simulation of high redshift LBGs and LAEs}
\label{sec:galfinder}
The cosmological simulation, the algorithm of galaxy finding and the
spectral modeling (UV continuum and \lya\ line) have been thoroughly described
in Paper I and II. Here we summarize the most relevant features
for this Paper. 

\subsection{The MareNostrum High-Z Universe Simulation}

The {\em MareNostrum  High-z Universe} simulation{\footnote{\tt http://astro.ft.uam.es/marenostrum}} follows the non linear evolution of structures in  
baryons (gas and stars) and dark matter, starting  from $z= 60$
 within a cube of $50\hMpc$ comoving on a side. 
The cosmological parameters used correspond to WMAP1 data
\citep{2003ApJS..148..175S} and are $\Omega_{\rm m}=0.3$, $\Omega_{\rm
  b}=0.045$, $\Omega_\Lambda=0.7$, $\sigma_8=0.9$, a Hubble parameter
$h=0.7$, and a spectral index $n=1$. The initial density field has
been sampled by $1024^3$ dark matter particles with a mass of $m_{\rm
  DM} = 8.2 \times 10^6 \hMsun $and $1024^3$ SPH gas particles with a
mass of $m_{\rm gas} = 1.4 \times 10^6 \hMsun$. The simulation has
been performed using the TREEPM+SPH code \texttt{GADGET-2}
\citep{Springel05}.   The gravitational smoothing scale  was set to 2
\hkpc in comoving coordinates.   We follow \cite{springel:03} to model the cooling, star formation and strong
kinetic feedback model in the form of galactic winds. We identify the objects  in the simulations  using the \texttt{AMIGA}
Halo Finder (\texttt{AHF}) \citep{Knollmann2009}. All objects with more than  $1000$ particles,  dark matter, gas and
stars combined, are used in our present analyses. We assume a  galaxy  is
resolved  if the object  contains   $200$ or more star particles, 
which  corresponds  to objects with $\gtrsim 400$
particles of gas.

\subsection{UV and \lya emission}

The photometric
properties of galaxies are calculated employing the stellar population
synthesis model STARDUST \citep{1999A&A...350..381D}. Using the methods
described in \cite{2003MNRAS.343...75H}. 

We consider only the  intrinsic \lya\ emission associated to star
formation.  We assume that the number of Hydrogen ionizing photons per
unit time is $1.8\times 10^{53}$ photons s$^{-1}$ for a star formation
rate of $1$ \Msun/yr \citep{1999ApJS..123....3L}. 

Assuming that $2/3$ of these photons
are converted to \lya photons (case-B recombination,
\citealt{1989agna.book.....O}), the intrinsic  \lya luminosity as a
function of the star formation rate is 
\begin{equation}
L_{Ly\alpha} = 1.9 \times 10^{42}\times ({\mathrm{SFR}}/{\mathrm
  M}_{\odot}\ {\mathrm{yr}}^{-1}) {\ \mathrm{erg\ s}}^{-1}.
\label{eq:lya_sfr}
\end{equation}
A change in the escape fraction of ionizing photons can thus induce changes in
the intrinsic \lya\ emission. In our model, we take this ionizing photon escape
fraction to be negligible. 

\subsection{Dust attenuation} 

Our approach to calculate the dust extinction is purely
phenomenological. The extinction curve for each galaxy is different depending on its
metallicity and gas contents \citep{1987A&A...186....1G}. The dust attenuation
model parametrizes both the extinction in a homogeneous ISM and in the
molecular clouds around young stars, following the physical model of
\cite{2000ApJ...539..718C}. The attenuation from dust in the homogeneous ISM
assumes a slab geometry, while the additional attenuation for young stars is
modeled using spherical symmetry.   

We first describe the optical depth for the homogeneous interstellar
medium, denoted by $\tau_{d}^{ISM}(\lambda)$. We take the mean
optical depth of a galactic disc at wavelength $\lambda$
to be  

\begin{equation}
\tau_{d}^{ISM}(\lambda)  = \eta
\left(\frac{A_{\lambda}}{A_{V}}\right)_{Z_{\odot}}\left(\frac{Z_g}{Z_{\odot}}\right)^r\left(\frac{\avg{N_{H}}}{2.1
  \times 10^{21} \mathrm{atoms\ cm}^{-2}}\right),
\label{eq:ISM}
\end{equation}
where $A_\lambda/A_V$ is the extinction curve from
\cite{1983A&A...128..212M}, $Z_{g}$ is the gas metallicity,
$\avg{N_H}$ is the mean atomic hydrogen column density and
$\eta=(1+z)^{-\alpha}$ is a factor that takes into account the
evolution of the dust to gas ratio at different redshifts.  

The extinction curve depends
on the gas metallicity $Z_{g}$ and is based on an interpolation
between the solar neighborhood and the Large and Small Magellanic
Clouds ($r=1.35$ for $\lambda < 2000 $\AA\ and $r=1.6$ for $\lambda >
2000$\AA).  

Stars
 younger than a given age, $t_{c}$, are subject to an additional attenuation
 in the birth clouds (BC) with optical depth 

\begin{equation}
\tau_{d}^{BC}(\lambda) = \left(\frac{1}{\mu}-1\right) \tau_{d}^{ISM}(\lambda),
\label{eq:young}
\end{equation}
where $\mu$ is the fraction of the total optical depth for these  young
stars with respect to that is found in the homogeneous ISM.

\subsection{\lya escape fraction}

In Paper II using our radiative transfer code {\tt CLARA} we obtained the \lya\
escape fraction for the corresponding slab described as a function of the
product $(a\tau_0)^{1/3}\tau_a$, where $\tau_0$ is the Hydrogen optical depth,
$\tau_a$ is the optical depth of absorbing material (for albedo values of $A$,
$\tau_{a}= (1-A)\tau_{d}$, where $\tau_{d}$ is the dust optical depth), and
$a$ is a measure of the gas temperature defined as
$a=\Delta\nu_L/(2\Delta\nu_D)$, $\Delta\nu_{D}=(v_{p}/c)\nu_0$ is the Doppler
frequency width, and $v_{p} = (2kT/m_{H})^{1/2}$ is $\sqrt{2}$ times the
velocity dispersion of the Hydrogen atom, $T$ is the gas temperature, $m_{H}$
is the Hydrogen atom mass and $\Delta\nu_L$ is the natural line width.  

We find that the equation

\begin{equation}
f_{\alpha} = \frac{1 - \exp(-P)}{P}
\label{eq:f_esc_homo}
\end{equation}
where
\begin{equation}
P = \epsilon((a\tau_{0})^{1/3}\tau_a)^{3/4},
\label{eq:f_esc_homo_bis}
\end{equation}
with $\epsilon = 3.5$ provides a reasonable
description of the Monte-Carlo results for $(a\tau_0)^{1/3}\tau_a <
200$ for the slab geometry with homogeneously distributed sources.

In addition to the foreground, homogeneous ISM extinction, we model
the attenuation due to the birth clouds. Under the physical conditions
for these poor neutral hydrogen clouds, $\tau^{BC}_{0}\sim \tau^{BC}_{a}$
together with $a \tau^{BC}_{0}<1$ and furthermore $\tau^{BC}_{a}>1$,
the enhancement in absorption by resonant scattering becomes irrelevant. Under these
conditions, we take the escape fraction as the continuum extinction at
$\lambda=1260$\AA\ for a spherical geometry. 

The effect of gas kinematics is not taken into account in the model. Appropriate outflow configurations can increase the \lya escape fraction \citep{1998A&A...334...11K,2006A&A...460..397V,2008A&A...488..491A}.

\subsection{Transmission through the IGM}

In Paper II we assumed that the IGM allows the transmission of 50 per cent of
the \lya line at every redshift $5\leq z\leq 7$. This is quantified by a
transmission coefficient ${\mathcal T}_{\rm IGM}=0.50$, which can be considered optimistic given recent numerical estimations.

Simplified analytic modeling \citep{stark2} of ${\mathcal T}_{\rm IGM}=\exp{(\tau_{\alpha})}$ where $\tau_{\alpha}$ is the optical depth for \lya photons (calculated from \cite{2006MNRAS.365..807M}), yield ${\mathcal T}_{\rm IGM}=0.58$ and ${\mathcal
  T}_{\rm IGM}=0.51$ at redshifts
$z\sim 5$ and $z\sim 6$. 3D Monte Carlo radiative transfer calculations of individual galaxies
  obtain values for the IGM transmission at $z\sim
  6$ of ${\mathcal T}_{\rm IGM}=0.26^{+0.13}_{-0.18}$ \citep{2011ApJ...728...52L}, a result that is readily explained by the effect of circumgalactic gas infall \citep{2007MNRAS.377.1175D,2008MNRAS.391...63I,2011MNRAS.410..830D}. 
Radiative transfer calculations by \cite{2010ApJ...716..574Z} find both a larger scatter and a lower mean value in the  fraction of intrinsic \lya radiation effectively detected at $z\sim 6$. However, in that calculation the authors need to  boost by a factor of $\sim 5$ the intrinsic \lya intensity in order to match the observational constraints from the luminosity function.

Our model requires a transmission fraction (${\mathcal T}_{\rm IGM}=0.5$) higher than the favored values for a symmetric \lya line (${\mathcal T}_{\rm IGM}< 0.4$). Due to the degeneracy between the escape fraction and the IGM transmission (the observed \lya luminosity is reduced by a factor $f_{\alpha}\times {\mathcal T}_{\rm IGM}$) lower transmission fractions ${\mathcal T}_{\rm IGM}$ can be compensated by higher escape fractions $f_{\alpha}$. Given that the clumpiness and amount of dust is already fixed by the LBG modelling, the freedom of changing $f_{\alpha}$ is restricted. The most probable interpretation is that outflows have to be considered in order to provide a physical picture that explains the values of ${\mathcal T}_{\rm IGM}\sim 0.5$ we need. As mentioned in the last section, outflows will also have an impact on the $f_{\alpha}$ escape fraction. Quantifying the effect of outflows is an issue we will addres in the future using the tools and simulations presented in Paper I and Paper II.

{At redshifts $z\gtrsim 6$, before reionization is completed, the estimations of the IGM transmission fraction depend on the assumptions made to model both the reionization and the \lya\ line radiative transfer \citep{2007MNRAS.381...75M,2011MNRAS.410..830D,2011MNRAS.414.2139D}. This makes even more difficult to reach a conclusive prediction on the expected values for ${\mathcal T}_{\rm IGM}$ at a fixed redshift and will influence the interpretation of our results beyond $z>6$, as will be discussed in Section 4.2.

\subsection{Comparison with the observational Luminosity Functions}

In Paper I we applied the above mentioned extinction model to match the rest-frame UV LFs
derived from the simulation to the observed LFs in the redshift range  $5\leq z\leq 7$.
The final model imposes additional extinction in all the stellar populations younger than $t_{c}=25$
Myr, with parameters $\mu=0.01$ for redshifts $z\sim 5,6$ and $\mu=0.03$ for redshift
$z\sim 7$, and $\alpha=1.5$. At the faint end ($M_{\rm
UV}<-20$) we have spotted a slight overabundance at all simulated redshifts
that could hint towards a low efficiency in the supernovae feedback that can
regulate the star formation process in the shallow potential wells of the
corresponding halos with masses $\sim 2\times 10^{10}$\hMsun.

In Paper II, with the dust contents already constrained from the UV LFs, we applied the
model for the \lya\ escape fraction to construct the LAE LFs. We note
  that the LAEs used to construct the LF had an equivalent width (EW) larger
  than a fiducial threshold of
$20$\AA, imposed to resemble a minimal threshold for
observational detection.  At redshifts $z\sim5$ and $z\sim 6$ we found a good match at the bright-end within the
Poissonian and cosmic variance uncertainties. The match with observations is also good if we
consider  IGM transmission fractions of 58 and 51 per cent as calculated by \cite{stark2} using the results by \cite{2006MNRAS.365..807M}.

However, there is an excess at the faint end of the simulated LF with
respect to the observations. The over-abundance corresponds to luminosities of
$L_{\ly} \sim 2\times 10^{42}$ erg/s, where we found as well a mild over-abundance at
the faint end of the UV LFs, if the star formation is lowered in some of these
numerical galaxies as to match the observed abundance, both the intrinsic UV
and \lya\ emission will drop, making the galaxies to fall below the detectable
range. Therefore, this change will not bear strong consequences for the
fraction $x_{\ly}$ because of the simultaneous changes in LBGs and LAEs abundance. 

At redshift $z\sim 7$ the normalization of the LAE LF is
still higher than observed at all luminosities by a factor of $\sim 0.4$ in
luminosity. By dimming each LAE in the simulation by a factor of $f_{\rm T}
= 0.4$ we can reproduce the  LF abundances at $z\sim 6.5$ \citep{2006ApJ...648....7K}.

\section{Results}
\label{sec:results}

\begin{figure}
\begin{center}
\includegraphics[width=0.45\textwidth,angle=270]{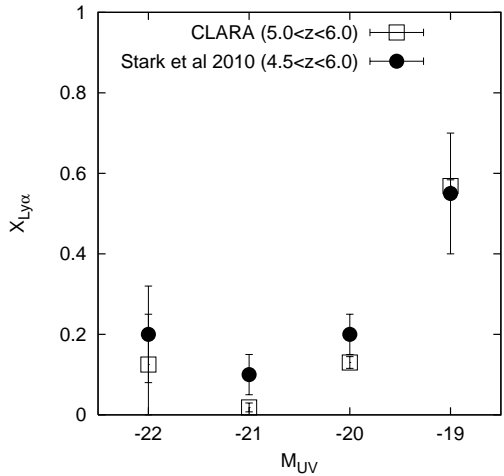}

\end{center}
\caption{
  Fraction of galaxies showing strong \lya\ emission as a function of the
  absolute $M_{\rm UV}$ magnitude in the models (empty symbols) and
  observations (filled circles). The error bars in the model are calculated
  from the Poissonian variance on the number of strong emitting LAEs. The results of our model (squares) in
  the redshift range $5\leq z\leq 6$ shows a broad agreement with the observations reported by
  \citet{stark1,stark2}.
\label{fig:fraction_base}}
\end{figure}

\begin{figure}
\begin{center}
\includegraphics[width=0.45\textwidth,angle=270]{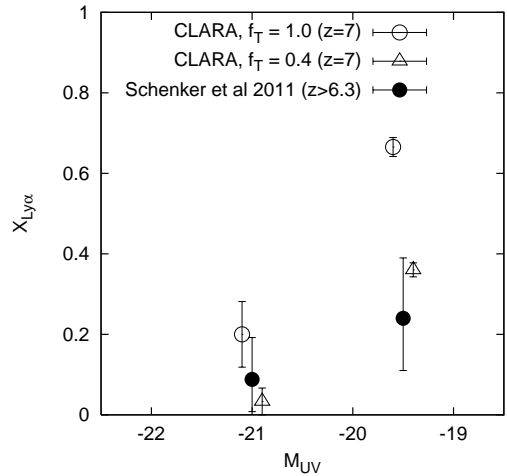}
\end{center}
\caption{
  Same as Figure \ref{fig:fraction_base}. Filled symbols represent the
  observational results by Schenker et al. (2011) for LBGs at $z>6.3$ with \lya
  emission with EW$>25$\AA. The empty symbols represent our
  results from the simulation at $z\sim 7$. The theoretical data-points have
  been shifted by $\pm0.1$ in redshift for clarity. The first scenario
  (circles) corresponds to the results of our model that do not include any
  evolution in the absorption model at $z\sim 7$ ($f_{\rm T}=1.0$), while the second (triangles)
  adds a further dimming each of LAE by a factor of $f_{\rm T}=0.4$. The factor
  was chosen to bring our results into agreement with the observed LAE LF at
  $z\sim 6.5$ \citep{2006ApJ...648....7K}. This scenario provides an improved match with the results reported
  by Schenker et al. (2011).
\label{fig:fraction_base_z7}}

\end{figure}

\subsection{Lyman-$\alpha$ fraction at $5<z<7$}

For each well resolved galaxy in the three simulation snapshots at $z=
5,6,7$ we calculate the UV magnitudes corrected by dust absorption, $M_{\rm
  UV}$, the slopes of the spectra between $1200$ \AA\ and $1600$ \AA, $\beta$,
together with the intrinsic and observed \lya\ luminosities.  Then, for each
galaxy we calculate the continuum flux red-wards of the \lya line at
$1240$\AA\ from the $\beta$ slope values and the flux calculated from the
$M_{\rm UV}$ magnitudes. This way of calculating  the continuum was chosen to
mimic the method used in \cite{stark1}. The \lya EW is then
calculated as the ratio of the observed intensity in the \lya line and the UV
continuum. 

We then bin the galaxies in $M_{\rm UV}$
and calculate the fraction of LBGs with \lya EW larger than $50$ \AA. In Fig
\ref{fig:fraction_base} we plot the fraction of galaxies with strong \lya\
emission ($x_{\ly}$) at redshifts $z=5$ and $z=6$ combined, and compared to the observational results at
$4.5<z<6.0$. As can be clearly seen in the plot, we are able to reproduce the
observational trend without any fine-tuning of our simulated galaxies. Namely
the values of $x_{\ly}$ for the bright LBGs with $M_{\rm UV}<-20.5$ and the increase
in the fraction of LAEs for faint LBGs with $M_{\rm
  UV}>-20.5$. The upturn in $x_{\ly}$ around $M_{UV}=-21.0$ is also
reproduced, which is a consequence of reproducing the shape of the UV LF which
shows the exponential drop in the abundance of LBGs beyond this magnitude,
making the fraction $x_{\ly}$ at the brightest bin sensitive to rise at large
values even for a low number of detected LBGs with strong \lya\ emission. On
average, a fraction $x_{\ly}=0.47\pm 0.01$ of the galaxies between $-20.5<M_{\rm UV}<-18.5$
show strong \lya\ emission with EW$>50$\AA, where only statistical
(poissonian) errors have been considered in this estimate.

\subsection{Evolution at $z>6$}

Recently, results of new line
  searches with deep Keck spectroscopy of $19$ LBGs in the redshift range
  $6.3<z<8.8$ selected with WFC3/IR data have been published \citep{2011arXiv1107.1261S}. The spectroscopic exposures were designed to reach EW
lower than 50\AA. Under these conditions, only 2 LAEs are convincingly
  detected. These results suggest a strong evolution with respect to the
  observations at $5<z<6$. Using complementary observational results from
  ultradeep optical spectroscopy with FORS2  on the Very Large Telescope
  (VLT) \citep{2010ApJ...725L.205F}, together with the
  results already obtained at $z\sim 6$ \citep{stark2}, the authors
  derive a fraction of LAEs with EW$>25$\AA: $x_{\ly}=0.088^{+0.088}_{-0.74}$
  ($-21.75<M_{\rm UV}<-20.25$) and $x_{\ly}=0.24\pm 0.15$
  ($-20.25<M_{\rm UV}<18.75$). It is important to keep in mind that these
  fractions (estimated from the measurements at $6.3<z<8.8$) are based
  on an extrapolation of the EW distributions at redshift $z=6$.

In Fig. \ref{fig:fraction_base_z7} we present the results of our model for the
  $x_{\ly}$ fraction (EW$>25$\AA) from the calculations with $f_{\rm T}=1.0$
  (i.e. no evolution from the $z=6$ extinction model) and
  $f_{\rm T}=0.4$.  We recall (see the final paragraph of \S 2.6 ) that the
  factor $f_{\rm T}$ is introduced   to tune the spatial abundance of LAEs at
  $z=7$ with respect to the LF   inferred from observations  
  
  Both models ($f_{\rm T}=1$ and $f_{\rm T}=0.4$) yield a similar trend with $M_{\rm UV}$ as the one found in
  $5\leq z\leq 6$. The fraction of LBGs with large EW increases for LBGs. However, the model
  that matches the LAE LF at $z\sim 7$ with $f_{\rm T}=0.4$ provides results for the
  $x_{\ly}$ fraction closer the the results derived from observations by
  \cite{2011arXiv1107.1261S} with $x_{\ly}=0.03 \pm 0.03$ ($-21.75<M_{\rm UV}<-20.25$) and
  $x_{\ly}=0.37\pm0.01$ ($-20.25<M_{\rm UV}<18.75$).

\begin{figure*}
\begin{center}
\includegraphics[width=0.42\textwidth,angle=270]{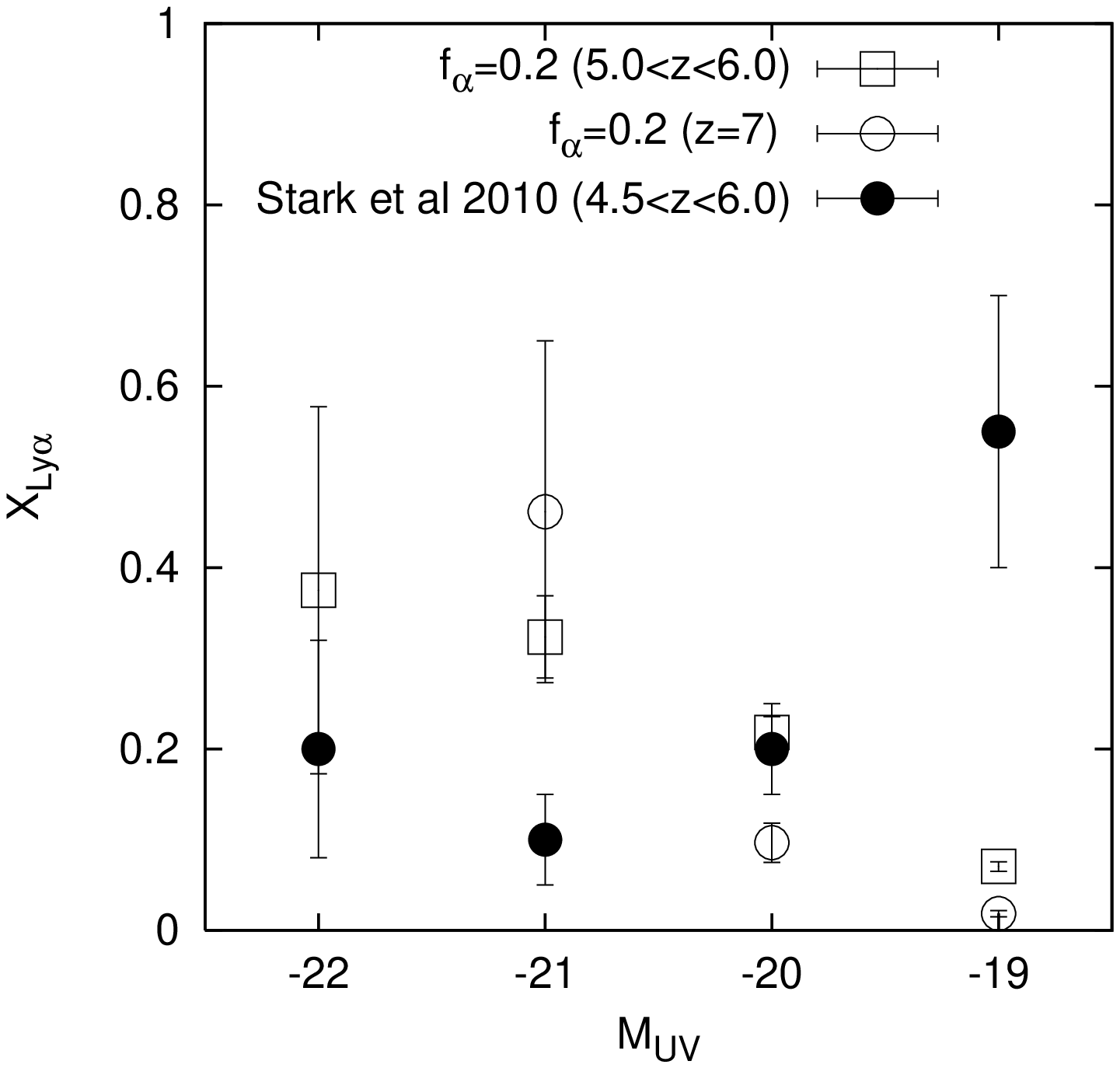}
\includegraphics[width=0.42\textwidth,angle=270]{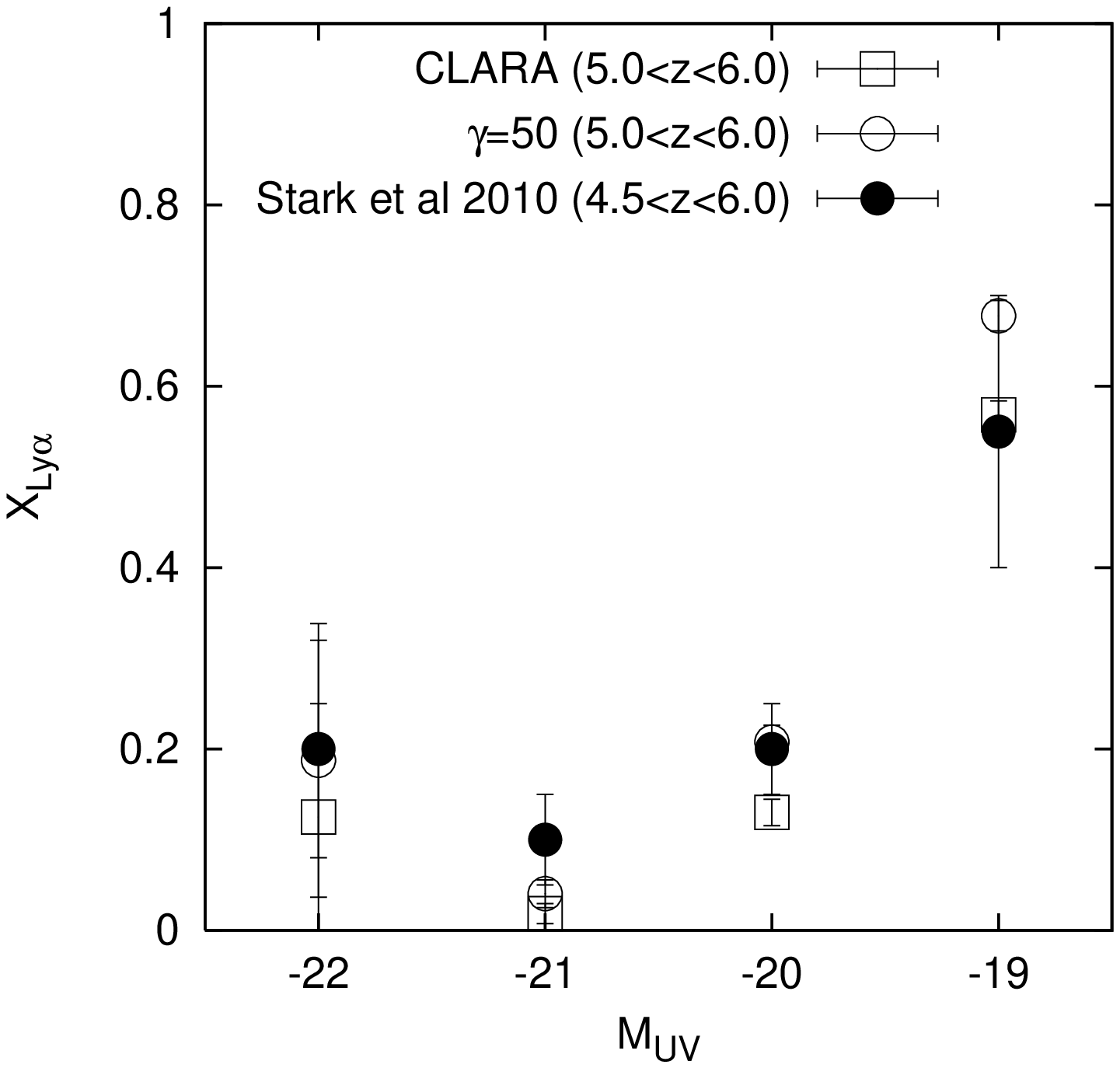}
\end{center}
\caption{
Same as Fig. \ref{fig:fraction_base} for two different models of the \lya\
  escape fraction. In both cases we keep the results for the $M_{\rm UV}$
  magnitudes fixed and vary the \lya escape fraction. The upper
  panel shows the results of a constant escape fraction $f_{\alpha}=0.2$,
  which clearly present a trend in contradiction with the observational constraints from
  \citet{stark1}. The lower panel shows the results of the
  model based on radiative transfer results (squares) compared against a phenomenological model
  where the \lya\ escape fraction is calculated by Eq.\ref{eq:pheno} (circles). The
  phenomenological approach shows a trend similar to the one obtained the
  radiative transfer model.}
\label{fig:fraction_pheno}
\end{figure*}

\section{Discussion}
\label{sec:discussion}
In the previous sections we have presented a model for LBGs and LAEs that
reproduce the observed trends for the $x_{\ly}$ fraction in two important
aspects: 1) the rise of $x_{\ly}$ for fainter $M_{\rm UV}$ magnitude at a fixed
redshift and 2) the overall evolution as a function of redshift between
$5\leq z\leq 7$.   In this Section we discuss the key elements in our model
behind these results.

\subsection{Dependence of $x_{\ly}$ on M$_{\rm UV}$}

  Assuming that the UV continuum and the \lya come from star formation,
  one can express the intrinsic luminosities as a function of the star
  formation rate: $L_{\lambda {\rm UV}}=1.4\times 10^{40} \times {\rm SFR}/(\Msun {\rm
  yr^-1})\ {\rm erg}\ {\rm s}^{-1}\ {\rm \AA}^{-1}$ and $L_{\ly} = 1.9\times
  10^{42} \times {\rm SFR/ \Msun {\rm yr}^{-1}}\ {\rm erg}\ {\rm
  s}^{-1}$   \citep{1998ARA&A..36..189K}. 
The exact conversion factors depend on the IMF and the metallicity of the
  stellar population. We will take these values as constant, although this approximation is not fundamental in the argument that follows.  

  Under these assumptions, the intrinsic equivalent width (EW$_{\rm i}$) is
  constant for all galaxies, EW$_{i}\sim 135$\AA, making all bright LBGs detectable as strong LAEs
  regardless of $M_{\rm UV}$ if one applies the EW cut $>50$\AA. The fact that
  only a fraction of bright LBG are also strong LAEs can be understood as a variation of the observed
  EW values from the intrinsic EW$_i$. This is naturally expected in the presence of extinction and/or preferential scattering of \lya photons out of the line of sight. 

  We can assume that extinction reduces the intensity of the line by a factor $f_{\alpha}$ and
  the UV continuum by a factor $f_c$. Further dimming by the neutral IGM can
  also reduce the \lya line intensity by a factor ${\mathcal T}_{\rm IGM}$. In this case, the measured equivalent
  width is EW$={\mathcal T}_{\rm IGM}\times (f_{\alpha }/f_{\rm c}) \times$EW$_{\rm
  i}$. In our model we have taken ${\mathcal T}_{\rm IGM}$ to be constant for all galaxies at
  a given redshift. Under this assumption the properties of the EW distribution can be attributed to the scatter in the
  extinction factors $f_{\rm c}$ and $f_{\alpha}$ \footnote{Although extreme values of the EW can be used to infer
  the presence of unusual stellar populations \citep{2010MNRAS.401.2343D}}.

Observationally, the factor $f_{c}$ can be inferred from the UV  spectral slope,
$\beta$. In the observed LBGs at $z>5$ \citep{2010ApJ...708L..69B,2011arXiv1102.5005D} and in our model as well, the evolution of the mean $\beta$ slope at a given magnitude $M_{\rm UV}$ is not strong, despite the presence of considerable scatter. On the other hand, the factor $f_{\alpha}$ can vary at least by a factor of 10 depending
on the properties of the galaxy \citep{CLARA}. This suggests that, at a fixed
redshift, the key physical factor in our model accounting for the trend of $x_{\ly}$ with $M_{\rm
  UV}$ is the dependence of the \lya escape fraction on galaxy luminosity.

To test this assumption, we show in the left panel of
Fig. \ref{fig:fraction_pheno} the results of approximating the \lya\ escape
fraction, $f_{\alpha}$, as a constant value. To construct this figure we keep
the results for the UV continuum fixed and reduce the \lya\ emission for all
galaxies by a constant value $f_{\alpha}=0.2$, a choice that also
  brings the LF for LAEs into agreement with observations at $z\sim 5$ and
  $z\sim 6$. This model fails to reproduce the observational trends for
$x_{\ly}$, in particular it seriously underestimates the fraction of bright
LAEs for faint LBGs to be $x_{\ly}=0.10\pm 0.01$. The \lya escape fraction is too low
for faint LBG, reducing too much the equivalent width and making most of these
galaxies undetectable as LAEs.

To further illustrate our point, we have also implemented a phenomenological model where the $f_{\alpha}$
escape fraction is calculated from an effective optical depth to \lya\ line
photons, which is assumed to be proportional to the continuum optical depth,

\begin{equation}
f_{\alpha} = \frac{1-\exp(-\gamma \tau_{d})}{\gamma\tau_{d}},
\label{eq:pheno}
\end{equation}
where $\gamma>1$ and $\tau_{d}$ is the continuum optical depth 
at a wavelength of $1216$\AA. \cite{2011arXiv1102.1509S} have
recently applied this model to reproduce the properties of LAEs at $z=3.1$. In
Fig. \ref{fig:fraction_pheno} we show the results for $\gamma=50$, this value
was chosen to give a close match to the LAE LF at $z\sim 5$ and $z \sim
6$. This model includes a dependence of the $f_{\alpha}$ escape fraction on galaxy gaseous mass (and hence luminosity), via the dependence on the gas
optical depth that is larger for luminous systems.  

This approximation manages to reproduce similar results for $x_{\ly}$ as our radiative
transfer motivated model, a result that is not surprising given the fact that
the functional form of Eq.(\ref{eq:pheno}) is close to the expected from
radiative transfer effects of the \lya\ line in the presence of neutral
Hydrogen \citep{2006MNRAS.367..979H,CLARA}. 

The result from this phenomenological model adds support to our claim that the
key factor in our model behind the shape of the $x_{\ly}$-$M_{\rm UV}$ plot is the mass
dependence of the \lya escape fraction that makes luminous galaxies have, on
average, small values $f_{\alpha}<0.2$ while fainter galaxies have large
$f_{\alpha}>0.2$.

\subsection{The evolution of $x_{\ly}$ beyond $z> 6$}

Different authors report a change in the properties of LAE emitting
galaxies beyond redshift $z>6$, both in the LF (\cite{2011ApJ...730....8H} and references therein) and the $x_{\ly}$
fraction \citep{2011arXiv1107.1261S}. 

In our model, the drop in the LAE abundance at $z\sim 6.5$ \citep{2006ApJ...648....7K} and the evolution of $x_{\ly}$ at $z>6.3$ \citep{2011arXiv1107.1261S} are explained by the evolution in the IGM transmission fraction. Such evolution in the transfer of the \lya line is required in our model to match the observational constraints. The dimming of all LAEs at $z\sim 7$ by a constant factor $f_{\rm   T}=0.4$ makes the results of our model to better fit the LAE LF at $z\sim 6.5$ and the $x_{\ly}$ fraction at $z>6.3$.

If confirmed, the observational evidence for a decrease in the observed $x_{\ly}$ fraction
  at redshifts $z>6$ can be attributed to three reasons: \begin{enumerate} \item[1)] an increase in the extinction in galaxies ISM
\item[2)] evolution in the Hydrogen neutral fraction in the IGM \item[3)] an increase in the
escape fraction of ionizing photons. 
\end{enumerate}

Given the observational and theoretical results for UV continuum slopes for
$z>6$ \citep{2010ApJ...708L..69B,2011arXiv1102.5005D,UVletter} it is rather
implausible that the extinction has increased. Therefore, only the explanations 
2) and 3) seem more plausible. It is beyond  scope of this paper to distinguish between
these two scenarios. 

However, if we add up the fraction $f_{\rm T}$ to the full transmission effect through a neutral
IGM we have ${\mathcal T}_{\rm IGM}(z=7.0)= f_{\rm T}\times 0.5 = 0.2$, where $0.5$ corresponds to
the fiducial value for the IGM transmission used in our model. The
interpretation of this fraction ${\mathcal T}_{\rm IGM}=0.2$ in terms of a global neutral
fraction $X_{HI}$ is considerably more difficult. Different models can give results in the
range $X_{HI}\sim 0.4-0.9$ depending on the treatment of the line through the
IGM, making it difficult to rule out large values of the neutral HI fraction from
the evolution in $x_{\ly}$, specially when gas bulk velocities from winds are
included in the analysis \citep{2007MNRAS.381...75M,2011MNRAS.414.2139D}.

\section{Conclusions}
\label{sec:discuss}
 
Observations report  that not all bright Lyman Break Galaxies in
the redshift range $5\leq z\leq 7$ are seen as strong Lyman Alpha Emitters 
\citep{stark1,stark2,2010ApJ...725L.205F,2011arXiv1107.1261S}. The fraction of LBGs that can be detected with strong
\lya\ emission, $x_{\ly}$ tests the equivalent width distribution, providing additional information that is not
included in the luminosity function of LAEs and LBGs.  
In this paper we present a theoretical model of high redshift LBGs and LAEs that reproduce the observed
trends for $x_{\ly}$ in the redshift range $5\leq z\leq7$. We find that the
a decreasing \lya escape fraction with increasing galaxy luminosity is a key element in our model to explain the observations.

In the redshift range $5\leq z\leq 6$ observations show a clear evolution of the $x_{\ly}$ with 
absolute magnitude $M_{\rm UV}$. From the results of our model, we suggest that a key ingredient to explain
this trend is the decrease of the \lya escape fraction with increasing galaxy
luminosity. We test this hypothesis by fixing the UV continuum results and
assuming a constant value for $f_{\alpha}$. This produces an
increasing fraction of LBGs with strong \lya\ emission for brighter $M_{\rm
  UV}$, the exact opposite trend of what is reported by observations. In
contrast, a model with a decreasing \lya\ escape fraction with increasing galaxy mass can simultaneously reproduce the observed trends for $x_{\ly}$ and the LAE-LF. The additional influence of an IGM transmission coefficient, ${\mathcal T}_{\rm IGM}$, dependent on large scale environment \citep{2011ApJ...726...38Z} is a point that deserves further investigation. In particular, it would be useful to know to what extent this kind of environment dependent transmission explains the trend of $x_{\ly}$ with absolute magnitude $M_{\rm UV}$.

In order to reproduce the $x_{\ly}$ evolution at $z>6.3$ our model requires
a decrease of the transmission fraction of \lya photons by a
factor of $f_{T}=0.4$, also improving the agreement with the observed LAE LF at $z\sim 6.5$, although confirmation of such evolution from increased sensitivity observations and larger samples is still needed. 

The fraction $x_{\ly}$ provides an additional statistical tool
to describe galaxy populations evolving during the reionization
epoch. More sensitive instruments and larger galaxy samples will be
necessary to quantify the evolution of the $x_{\ly}$
fraction. In the same spirit, this statistic can be used as a benchmark for a theoretical model aiming at describing the evolving galaxy populations at these redshifts.

\section*{Acknowledgments}
We thank the anonymous referee for the careful reading and the thoughtful
comments which helped us to improve the clarity of the paper. We thank Matthew Schenker for providing us with the results from his paper in
electronic format, and for useful clarifications on the procedure to calculate
the $x_{\ly}$ fraction.

The simulation used in this work is part of the MareNostrum Numerical
Cosmology Project at the BSC. The data analysis has been performed at the NIC
Juelich and the LRZ Munich.

JEFT and FP acknowledge the support by the ESF ASTROSIM network through the
short visit grant scheme that helped in the development process of {\tt
  CLARA}.

GY acknowledges support of MICINN (Spain) through research grants
FPA2009-08958 and AYA2009-13875-C03-02. 
We equally acknowledge funding from the Consolider project MULTIDARK (CSD
2009-00064) and the Comunidad de Madrid project ASTROMADRID (S2009/ESP-146).

\bibliographystyle{mn2e}

\bibliography{references}

\end{document}